\documentclass[prd,aps,twocolumn,preprintnumbers,amsmath,amssymb]{revtex4}
\usepackage{amsmath}
\usepackage{graphicx}
\usepackage{dcolumn}
\usepackage{bm}
\usepackage{amssymb}
\usepackage{latexsym}
\usepackage{hhline}
\usepackage{multirow}

\def\be{\begin{equation}}
\def\ee{\end{equation}}
\def\bea{\begin{eqnarray}}
\def\eea{\end{eqnarray}}

\begin{document}

\title{Reconstruction of $f(R)$ models with Scale-invariant Power Spectrum}

\author{Taotao Qiu\footnote{xsjqiu@gmail.com,qiutt@ntu.edu.tw}}

\affiliation{$1$. Leung Center for Cosmology and Particle Astrophysics National Taiwan University, Taipei 106, Taiwan}
\affiliation{$2$. Department of Physics, National Taiwan University, Taipei 10617, Taiwan}

\begin{abstract}
Following our previous work in [JCAP {\bf 1206}, 041 (2012) \cite{Qiu:2012ia}], in this paper, we continue our study of reconstructing $f(R)$ modified gravity models that can be connected to a single scalar field in general relativity via conformal transformation, which lead to scale-invariant power spectrum in the early universe. With $f(R)$ modified gravity, one does not need to introduce extra scalar, the nature of which are to be explained. Different from general nonminimal coupling theory, the behavior of the $f(R)$ theory has been fixed by its counterpart in Einstein frame, and thus have one to one correspondence. Numerical plots of the functional form of $f(R)$ as well as the evolution of $R$ in terms of cosmic time $t$ are also presented.
\end{abstract}
\maketitle

\section{Introduction}
For theories of the early universe, the right amount of perturbations must be generated so as to conform with our observations such as cosmic miscrowave background (CMB) \cite{Larson:2010gs} and large scale structures (LSS) \cite{Bernardeau:2001qr}. One of the well-known observed features is, the power spectrum of these perturbations, which comes from the 2-point correlation function, has to be (nearly) scale-invariant \cite{Larson:2010gs}, which will put on non-trivial constraints on theoretical model building. Although it is well-known that a single scalar field, which drives the universe into de-Sitter like expansion (inflation \cite{Starobinsky:1980te,Guth:1980zm,Albrecht:1982wi,Linde:1983gd,Starobinsky:1985ww}, while the scalar is called inflaton), or nonrelativistic matter-like contraction \cite{Finelli:2001sr,Cai:2007qw} could easily generate perturbations to meet the requirement, the nature of the scalar is still unclear.

Scale-invariant power spectrum may also arise when one modify Einstein's gravity at early times. In some cases, the modified gravity theories could be connected with unmodified general relativity (GR) plus a scalar through conformal transformations \cite{Faraoni:1998qx}, with the latter being viewed as the counterpart in Einstein frame of the former. Due to the equivalence between the two frames (Jordan and Einstein), the perturbation generated by the couple of counterparts are exactly the same. Thanks to the connection, one can thus reconstruct models of modified gravity from the known evolution of GR plus a scalar models, which can lead to inflation or matter-contraction scenarios. Recently we proposed a way of reconstructing the models with a scalar nonminimally coupled to gravity which could give rise to scale-invariant power spectrum \cite{Qiu:2012ia}. In this paper, we will consider another case of modified gravity, namely $f(R)$ theories. Actually as we will see later, $f(R)$ theories could be one specific but nontrivial form of nonminimal coupling. In $f(R)$ theories, there is no need to introduce the unknown scalar, and the universe is driven totally by its gravitational structure. $f(R)$ theories has been used widely as alternatives of inflation, dark matter, dark energy and so on. See \cite{Faraoni:2000gx} for comprehensive reviews.

The reconstruction of $f(R)$ gravity has been pursued by many authors, see \cite{Nojiri:2006be}. In their approaches, most of them reconstruct $f(R)$ theory in Jordan frame itself, provided that the cosmic evolution in Jordan frame is given. Here we will reconstruct in a different way, namely from their counterpart in Einstein frame, which looks like a single scalar field in GR, via conformal transformation. This kind of reconstruction aims at connecting different evolutions of the universe driven by modified gravity in its Jordan and Einstein frames. As is shown in \cite{Qiu:2012ia}, in Einstein frame there are only two cases which could give rise to (nearly) scale-invariant power spectrum, namely inflation and matter-contraction. Taking the Einstein frame lagrangian as: \be\label{einstein} {\cal L}_E\sim \frac{1}{2}R_E-\frac{1}{2}(\partial\varphi_E)^2-V(\varphi_E)~,\ee where here and after we set the unit such that $8\pi G=M_{Pl}^{-2}=1$, and use the metric signature $(-,+,+,+)$. A simple and representative solution is the exact solution which is obtained assuming that its equation of state $w_E$ is a constant, namely: \bea\label{parametrize} a_E(t_E)&\sim&(\pm t_E)^{\frac{2}{3(1+w_E)}}~,~H_E(t_E)=\frac{2}{3(1+w_E)t_E}~,\nonumber\\ \varphi_E(t_E)&=&\frac{2\ln(\pm M t_E)}{\sqrt{3(1+w_E)}}~,~V(\varphi_E)=V_0 e^{-\sqrt{3(1+w_E)}\varphi_E}~\eea where $M$ is some energy scale. In this parametrization, we have set $``+"$ for positive $t_E$ meaning an expanding phase, while $``-"$ for negative $t_E$ denoting a contracting phase, and $V_0$ is some constant factor. In Inflation case, we have $w_E=-1+2\epsilon_E/3$ with the slow-roll parameter $|\epsilon_E|\equiv|-(dH_E/dt_E)/H_E^2|\ll1$, then Eq. (\ref{parametrize}) can be written as: \bea\label{parametrizeinf} a_E(t_E)&\sim&{t_E}^{\frac{1}{\epsilon_E}}~,~H_E(t_E)=\frac{1}{\epsilon_E t_E}~,\nonumber\\ \varphi_E(t_E)&=&\sqrt{\frac{2}{\epsilon_E}}\ln(Mt_E)~,~V(\varphi_E)=V_0 e^{-\sqrt{2\epsilon_E}\varphi_E}~,\eea while in matter-contraction case, one has $w_E=0$, and thus Eq. becomes: \bea\label{parametrizeMB} a_E(t_E)&\sim& (-t_E)^{\frac{2}{3}}~,~H_E(t_E)=\frac{2}{3t_E}~,\nonumber\\ \varphi_E(t_E)&=&\frac{2}{\sqrt{3}}\ln(-Mt_E)~,~V(\varphi_E)=V_0 e^{-\sqrt{3}\varphi_E}~.\eea

In this short paper, we will mainly focus on the Jordan frame of the modified gravity theories in order to find which form can be conformally connected to the above two cases, while more complete study for the case of varying $w_E$ (or $\epsilon_E$) will be left for the future.

The remaining sections are organized as following: in Sec. II, we review the main results for the general nonminimal coupling theories that was obtained in our previous paper. in Sec. III, we focus on $f(R)$ theories. Numerical plots of the functional form of $f(R)$ as well as the evolution of $R$ in terms of cosmic time $t$ are presented. Furthermore, we also discussed about the relation of the evolutions of various cosmological variables between the two frames for an arbitrary constant $\epsilon_E$. In Sec. IV we conclude our paper.

\section{Review of reconstruction of nonminimal coupling theory}
\subsection{Background}
First of all, we will briefly review the main results obtained in \cite{Qiu:2012ia}. The action of the nonminimal coupling theory we are considering is: \be\label{actionNMC} {\cal S}_{NMC}=\int d^4x\sqrt{-g}\Bigl[F(\phi)R-\frac{1}{2}Z(\phi)\partial_\mu\phi\partial^\mu\phi-U(\phi)\Bigr]~,\ee where $F(\phi)$ and $Z(\phi)$ can be arbitrary functions of the field $\phi$ in the Jordan frame, and $U(\phi)$ is the potential. The equation of motion of $\phi$ is: \be\label{eomNMC} \ddot\phi+3H_J\dot\phi+\frac{Z_\phi}{2Z}\dot\phi^2-\frac{6F_\phi}{Z}(\dot H+2H^2)+\frac{U_\phi}{Z}=0~,\ee where subscript ``$\phi$" indicates $\partial/\partial\phi$ and dot denotes derivative with respect to cosmic time $t_J$ in the Jordan frame, and the Friedmann Equation is: \be\label{friedmannNMC} 6H_J\dot F+6H_J^2F=\frac{1}{2}Z\dot\phi^2+U~.\ee Following the conformal transformation of metrics in Jordan and Einstein frame, $g_{\mu\nu}^{(E)}=\Omega^2 g_{\mu\nu}^{(J)}$, where $\Omega^2\equiv 2F$, the relations of some basic variables between the two frames are summarized as follows: \bea\label{relationNMC} dt_E&=&\Omega dt_J~,~a_E=\Omega a_J~,~H_E=\frac{H_J}{\Omega}(1+\frac{\dot\Omega}{2H_J\Omega})~,\nonumber\\ \varphi_E&=&\int\sqrt{\frac{6M_{Pl}^2\Omega_\phi^2+Z}{\Omega^2}}d\phi~,~V(\varphi_E)=\frac{U(\phi)}{\Omega^4}~.\eea
\subsection{Perturbations}
The equation of motion of the perturbation generated by the action (\ref{actionNMC}) can be written down as: \be\label{perteomNMC} u^{\prime\prime}_{\cal R}+(k^2-\frac{(a_J\sqrt{2Q_{\cal R}})^{\prime\prime}}{a_J\sqrt{2Q_{\cal R}}})u_{\cal R}=0~,\ee where $u_{\cal R}=a_J\sqrt{2Q_{\cal R}}{\cal R}$, and ${\cal R}$ is the conformal-invariant curvature perturbation. The variable $Q_{\cal R}$ is defined as: \be Q_{\cal R}\equiv\frac{2F}{(2+\delta_F)^2}[3\delta_{F}^{2}+\frac{\dot\phi^2Z}{H_J^{2}F}]~,\ee where $\delta_F\equiv\dot F/(H_J F)$. The prime denotes derivative with respect to the conformal time $\eta=\int a_J^{-1}(t_J)dt_J$. With the parametrization that $a_J\sqrt{2Q_{\cal R}}\sim |\eta_\ast-\eta|^\lambda$, the superhorizon solution of Eq. (\ref{perteomNMC}) can be expressed in the following: \bea\label{resultNMC}
u_{\cal R}&\sim&\sqrt{|\eta_\ast-\eta|}\Big[c_1J_{\lambda-\frac{1}{2}}(k|\eta_\ast-\eta|)+c_2J_{\frac{1}{2}-\lambda}(k|\eta_\ast-\eta)\Big]\nonumber\\  &\sim& c_1k^{\lambda-\frac{1}{2}}|\eta_\ast-\eta|^{\lambda-\frac{1}{2}}+c_2k^{\frac{1}{2}-\lambda}|\eta_\ast-\eta|^{1-\lambda}~,\nonumber\\
{\cal R}&=&\frac{u_{\cal R}}{a_J\sqrt{2Q_{\cal R}}}\sim c_1k^{\lambda-\frac{1}{2}}+c_2k^{\frac{1}{2}-\lambda}|\eta_\ast-\eta|^{1-2\lambda}~,\eea where $J_i$ is the Bessel function and $c_1$, $c_2$ are constants. The power spectrum is defined as \be {\cal P}_{\cal R}(k)\equiv\frac{k^3}{2\pi^2}\big|{\cal R}\big|^2~.\ee

From the above solution, it is straightforward to see that scale-invariant spectrum $({\cal P}_{\cal R}(k)\sim k^0)$ can be obtained in two ways: one is $\lambda=-1$, where the time-varying mode becomes decaying while the constant mode dominates the perturbation, which is inflation, and the other is $\lambda=2$, where the time-varying mode is the growing mode and thus dominates over the constant one, which is matter-contraction. In fact, from the relation (\ref{relationNMC}) one can express $Q_{\cal R}$ as: \be Q_{\cal R}\sim F\epsilon_E~,\ee and since we have assumed constant $w_E$ and $\epsilon_E$, the condition of getting scale-invariant power-spectrum can be written as $a_J\sqrt{F}\sim|\eta_\ast-\eta|^{-1}$ or $a_J\sqrt{F}\sim|\eta_\ast-\eta|^2$.
\subsection{Reconstruction of nonminimal coupling theory in Jordan Frame}
We can reconstruct the universe evolution once we assume the evolution of $\Omega$ in terms of $t_J$. In our previous paper \cite{Qiu:2012ia}, we assumed that $\Omega(t_J)=\Omega_0 [(\pm t_J)/(\pm t_J^\ast)]^{\omega}$, then from the relation (\ref{relationNMC}) we have: \bea\label{tE2tJI1} t_E=\left\{ \begin{array}{l} \frac{\Omega_0t^\ast_J}{\omega+1}\Big(\frac{\pm t_J}{\pm t^\ast_J}\Big)^{\omega+1}~~~~{\rm for}~~\omega\neq-1~,\\\\ \Omega_0t^\ast_J\ln(\pm\bar{t}_J)~~~~{\rm for}~~\omega=-1~,\\ \end{array}\right. \eea where the $``+"$ sign in $``\pm"$ means $t_J>0$, and in the Jordan frame the universe is expanding, while the $``-"$ sign means $t_J<0$, and in the Jordan frame the universe is contracting. Here we define $\bar{t}_J=t_J/t_{Pl}$ where $t_{Pl}$ is the Planck time. Substituting it into Eqs. (\ref{parametrizeinf}) and (\ref{parametrizeMB}) respectively, one can get the evolution of variables such as $a_J$, $H_J$ and $w_J$ in terms of $t_J$ as (for $\omega\neq-1$ only):  \bea a_J(t_J)&\sim&(\pm t_J)^{\frac{1+(1-\epsilon_E)\omega}{\epsilon_E}}~,~H_J=\frac{1+(1-\epsilon_E)\omega}{\epsilon_Et_J}~,\nonumber\\ w_J&=&-1+\frac{2}{3}\frac{\epsilon_E}{1+(1-\epsilon_E)\omega}~,\eea where $|\epsilon_E|\ll1$ for the case corresponding to inflation, while $\epsilon_E=3(1+w_E)/2=3/2$ for the case corresponding to matter-contraction, respectively. Moreover, from relation (\ref{relationNMC}) one can also find the evolution of field variables, and thus determine the form of functions $F(\phi)$, $Z(\phi)$ and $U(\phi)$ in the lagrangian. In fact, taking the ansatz of $Z(\phi)=Z_0\phi^{2z}$ and $U(\phi)=U_0\phi^{q}$, and with the help of Eqs. (\ref{eomNMC}) and (\ref{friedmannNMC}), we found the relation: \be\label{relationfieldNMC} F(\phi)=F_0\phi^{2z+2}~, q=2(z+1)(1-\frac{1}{\omega})~,\ee and the equation of state $w_J$ can be given by: \be w_J=\frac{2(z+1)(5\epsilon_E-6)-q(2\epsilon_E-3)}{3[2(z+1)(2-\epsilon_E)-q]}~.\ee

From above we can see that, once the functional form of $F(\phi)$, $Z(\phi)$ and $U(\phi)$ in action (\ref{actionNMC}) is given by (\ref{relationfieldNMC}), one could obtain scale-invariant power spectrum. Rather than being fixed to be inflation or matter-contraction only, the evolution of the universe in the Jordan frame has more freedom. This is because in the Jordan frame, the nonminimal coupling action (\ref{actionNMC}) has more degrees of freedom than that in the Einstein frame and is more dependent on the form of the action. However, as we will see below, it is not the case in $f(R)$ theory. In $f(R)$ theory, there will be less degree of freedom than nonminimal coupling theory and the form of $f(R)$ will be more fixed. Following similar steps, we will find the appropriate $f(R)$ theory, which can correspond to inflation or matter-contraction scenarios in its Einstein frame and thus, give rise to scale-invariant power spectrum.
\section{Reconstruction of $f(R)$ modified gravity theory}
\subsection{Background}
Now we turn on to study the reconstruction of $f(R)$ modified gravity theories. The action of $f(R)$ modified gravity theory is:
\be\label{actionfr} {\cal S}_{f(R)}=\int d^{4}x\sqrt{-g}f(R)~,\ee where $f(R)$ can be arbitrary function of the Ricci scalar $R$. Varying the action (\ref{actionfr}) with respect to the metric $g_{\mu\nu}$ we can get the equation of motion: \be\label{eomfr} -F_{,\mu;\nu}+g_{\mu\nu}\Box F+FR_{\mu\nu}-\frac{1}{2}g_{\mu\nu}f=0~,\ee where we defined the function $F(R)\equiv \partial f/\partial R$. The left part of the above equation can also be viewed as the ``effective" stress energy tensor $\Sigma_{\mu\nu}$ of $f(R)$ modified gravity, which satisfies the continuity equation, $\nabla^\mu\Sigma_{\mu\nu}=0$. Moreover, the ``$0-0$" and ``$0-i$" components of Eq. (\ref{eomfr}) are just Friedmann equations, which are \be 3H^2F=\frac{1}{2}(f+3\ddot F+3H\dot F)~,~-2\dot H F=\ddot F-H\dot F~,\ee respectively.

Same as nonminimal coupling theory, $f(R)$ theories with action (\ref{actionfr}) can also be connected with (\ref{einstein}) as its counterpart in the Einstein frame, via the conformal transformation $g_{\mu\nu}^{(E)}=\Omega^2 g_{\mu\nu}^{(J)}$ with $\Omega^2=2F$. To see this, one can rewrite the action (\ref{actionfr}) in the form of scalar-tensor theory, namely as: \be\label{actionst} {\cal S}_{ST}=\int d^{4}x\sqrt{-g}\Big[F(R)R-U(R)\Big]~\ee where the potential $U(R)$ can be identified as $F(R)R-f(R)$. The relations of the basic variables between the two frames are summarized as follows: \bea\label{relationfr} dt_E&=&\Omega dt_J~,~a_E=\Omega a_J~,~H_E=\frac{H_J}{\Omega}(1+\frac{\dot\Omega}{2H_J\Omega})~,\nonumber\\ \varphi_E&=&\sqrt{6}\ln\Omega~,~V(\varphi_E)=\frac{U(R)}{\Omega^4}~.\eea

From the transformed action (\ref{actionst}) we can see that, the $f(R)$ action is actually the specific form of the general nonminimal coupling action (\ref{actionNMC}) with $Z(\phi)=0$, as long as we identify $F(\phi)$ with $F(R)$, and $U(\phi)$ with $U(R)$, which is easy provided that the inverse function of $F(\phi)$ exists. Moreover, since $Z(\phi)$ as well as the kinetic term of (\ref{actionNMC}) vanishes, there are less degrees of freedom in $f(R)$ than in nonminimal coupling theories, and the conformal factor $\Omega$, which determines the cosmic evolution in Jordan frame, can be totally fixed by the field $\varphi_E$. Therefore, when there is one kind of evolution in Einstein frame, there is only one kind of evolution in Jordan frame. This gives less possibilities for $f(R)$ theories to get scale-invariant power spectrum than those for nonminimal coupling theories.
\subsection{Perturbations}
One can also check from the perturbation theory of $f(R)$ that what conditions should be met when one requires a scale-invariant power spectrum. Working in the Arnowitt-Deser-Misner (ADM) formalism \cite{Arnowitt:1962hi}, one can obtain the perturbed action of $f(R)$ up to the second order as: \be\label{pertactionfr} {\cal
S}^{(2)}=\int d\eta d^3xa_J^2 Q_{\cal R}\Bigl[{\cal R}^{\prime 2}-(\partial{\cal R})^2\Bigr]~,\ee where ${\cal R}$ is the conformal-invariant curvature perturbation, and \be Q_{\cal R}\equiv \frac{6F\delta_{F}^{2}}{(2+\delta_F)^2}~\ee with $\delta_F=\dot F/(H_J F)$ and the prime denotes derivative with respect to the conformal time $\eta$. Varying (\ref{pertactionfr}) with respect to ${\cal R}$, one can straightforwardly write down the equation of motion for the perturbation as: \be\label{perteomfr} u^{\prime\prime}_{\cal R}+(k^2-\frac{(a_J\sqrt{2Q_{\cal R}})^{\prime\prime}}{a_J\sqrt{2Q_{\cal R}}})u_{\cal R}=0~,\ee through the redefined variables $u_{\cal R}=a_J\sqrt{2Q_{\cal R}}{\cal R}$.

From the above analysis, we can directly conclude that scale-invariant spectrum can be obtained in two ways, namely $a_J\sqrt{2Q_{\cal R}}\sim|\eta_\ast-\eta|^{-1}$ which corresponds to inflation, or $a_J\sqrt{2Q_{\cal R}}\sim|\eta_\ast-\eta|^2$ which corresponds to matter-contraction. Moreover, from the relation (\ref{relationfr}) one can express $Q_{\cal R}$ as $Q_{\cal R}\sim F\epsilon_E$, the same as that in nonminimal coupling theories. Here we can see again that $f(R)$ theories are nothing but specific case of nonminimal coupling theories. In our case where constant $w_E$ and $\epsilon_E$ have been assumed, the condition of getting scale-invariant power-spectrum can be written as $a_J\sqrt{F}\sim|\eta_\ast-\eta|^{-1}$ or $a_J\sqrt{F}\sim|\eta_\ast-\eta|^2$.
\subsection{Reconstruction of $f(R)$ modified gravity theory in Jordan Frame}
First of all, from relations (\ref{relationfr}) as well as the evolution of $\varphi_E(t_E)$ in the Einstein frame (\ref{parametrize}), we can obtain the evolution of the conformal factor $\Omega$ in terms of $t_E$, which is \be \Omega=\Big(\frac{t_E}{t_E^\ast}\Big)^{\frac{1}{\sqrt{3\epsilon_E}}}~,~|\epsilon_E|\ll1~,\ee where $t_E^\ast=M^{-1}$. Since the universe in Einstein frame is expanding, we set $t_E$ and $t_E^\ast$ to be positive \footnote{Here and after, we assume that the same as $t_E$, $t_J$ monotonically increases, although its value can be either positive or negative. This is an arbitrary choice, only indicating the arrow of time, and one can surely assume that time goes in an opposite direction, which is only trivially dual to the current case by the transformation $t_J^\prime\rightarrow-t_J$.}. Since $dt_J=\Omega^{-1}(t_E)dt_E$, one could easily get $t_J$ as: \be t_J=\frac{\sqrt{3\epsilon_E}t_E^\ast}{\sqrt{3\epsilon_E}-1}\Big(\frac{t_E}{t_E^\ast}\Big)^{\frac{\sqrt{3\epsilon_E}-1}{\sqrt{3\epsilon_E}}}~,\ee or equivalently, \be \frac{t_E}{t_E^\ast}=\Big(\frac{-t_J}{-t_J^\ast}\Big)^{\frac{\sqrt{3\epsilon_E}}{\sqrt{3\epsilon_E}-1}}~,~t_J^\ast\equiv\frac{\sqrt{3\epsilon_E}t_E^\ast}{\sqrt{3\epsilon_E}-1}~.\ee Note that since $|\epsilon_E|\ll1$, $t_J$ and $t_J^\ast<0$. Then we have: \be\label{omegatj} \Omega(t_J)=\Big(\frac{-t_J}{-t_J^\ast}\Big)^{\frac{1}{\sqrt{3\epsilon_E}-1}}~.\ee

With Eqs. (\ref{parametrizeinf}), (\ref{relationfr}) and (\ref{omegatj}) in hand, we can obtain the evolution of $a_J$, $H_J$ and $w_J$ in the Jordan frame, in terms of $t_J$. The results are: \bea \label{evolutioninf} a_J(t_J)&\sim&\Big(\frac{-t_J}{-t_J^\ast}\Big)^{\frac{\sqrt{3}-\sqrt{\epsilon_E}}{\sqrt{\epsilon_E}(\sqrt{3\epsilon_E}-1)}}~,\nonumber\\ H_J(t_J)&=&\frac{\sqrt{3}-\sqrt{\epsilon_E}}{\sqrt{\epsilon_E}(\sqrt{3\epsilon_E}-1)t_J}~,\nonumber\\ w_J&=&\frac{\sqrt{\epsilon_E}+2\sqrt{3}\epsilon_E-3\sqrt{3}}{3(\sqrt{3}-\sqrt{\epsilon_E})}~. \eea

From this result we can see that, since $|\epsilon_E|\ll1$ as we considered, the index of $a_J$ in terms of $t_J$ (namely $1/\epsilon_J$, if we define $\epsilon_J$ to be the slow-roll parameter in the Jordan frame) is less than zero, and $a_J(t_J)$ will be increasing as $t_J$ increases. This indicates that it is an expanding universe, driven by $f(R)$ modified gravity theory, which is equivalent to the so-called ``Super-inflation"  \cite{Gunzig:2000kk} (or phantom-inflation \cite{Piao:2003ty}) scenario in GR when transformed to the Einstein frame. One can also look into the equation of state $w_J$ of the universe, which is very much close to $-1$ up to order of slow-roll parameter, which means that the universe in the Jordan frame is also near de Sitter, so different from the general nonminimal coupling theory, inflation in the Einstein frame can only refer to inflation in the Jordan frame in $f(R)$ modified gravity theory.

The Ricci scalar $R$, which is defined as $R=6(\dot H+2H^2)$, can be expressed as: \be\label{ricciinf} R(t_J)=6\frac{(2-\sqrt{3\epsilon_E})(3-\epsilon_E)}{\epsilon_E(1-\sqrt{3\epsilon_E})^2t_J^2}~.\ee Finally, with Eqs. (\ref{omegatj}), (\ref{ricciinf}), as well as the relation $\Omega^2=2F$, we can obtain the form of $F(R)$ as: \be F(R)=\frac{1}{2}\Big(\frac{R}{R_0^{inf}}\Big)^{\frac{1}{1-\sqrt{3\epsilon_E}}}~,~R_0^{inf}\equiv6\frac{(2-\sqrt{3\epsilon_E})(3-\epsilon_E)}{\epsilon_E(1-\sqrt{3\epsilon_E})^2{t_J^\ast}^2}~\ee and \bea\label{frinf} f(R)&=&\int F(R)dR~\nonumber\\ &=&\frac{1-\sqrt{3\epsilon_E}}{4-2\sqrt{3\epsilon_E}}R_0^{inf}\Big(\frac{R}{R_0^{inf}}\Big)^{\frac{2-\sqrt{3\epsilon_E}}{1-\sqrt{3\epsilon_E}}}~.\eea We can see that when $\epsilon_E$ is small during inflation, the function of $f(R)$ is almost proportional to $R^2$ up to slow-roll parameter. Therefore, this model coincides with the well-known Starobinsky's model \cite{Starobinsky:1980te} of which $f(R)\sim R+\alpha R^2$ in the very early time, when $R$ is very large. In the late time when $\epsilon_E$ is large, it goes near the standard GR.

\begin{figure}[htbp]
\centering
\includegraphics[scale=0.5]{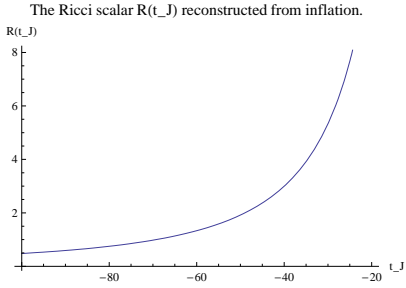}
\caption{The behavior of $R(t_J)$ w.r.t. $t_J$, where we choose $M=0.1$ and hence $t_E^\ast=10$. In this case, $R>0$, and is increasing w.r.t. $t_J$, showing a ``super/phantom-inflation" behavior.}\label{Ricci1plot}
\end{figure}

\begin{figure}[htbp]
\centering
\includegraphics[scale=0.5]{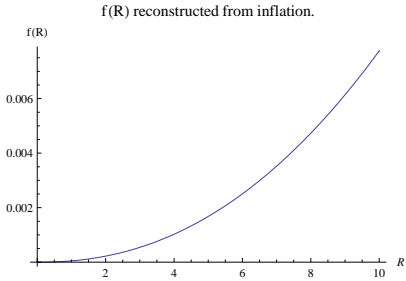}
\caption{The behavior of $f(R)$ w.r.t. $R$, where we choose $M=0.1$ and hence $t_E^\ast=10$. We can see that $f(R)$ monotonically increases with $R$, and in the limit of large $R$, it approaches to the squared power-law $f(R)\sim R^2$.}\label{f1}
\end{figure}

The plot of $R(t_J)$ and $f(R)$, which are reconstructed from inflation in its Einstein frame, are presented in Figs. \ref{Ricci1plot} and \ref{f1}.


Following the same procedure, we can do the reconstruction of $f(R)$ from matter contraction, just replacing $t_E$ by $-t_E$, and $\epsilon_E$ by the value $3/2$. Note that here $t_E$ and $t_E^\ast$ are negative. $t_J$ and $\Omega(t_J)$ will become \be t_J=t_J^\ast\Big(\frac{-t_E}{-t_E^\ast}\Big)^{1-\frac{\sqrt{2}}{3}}~,t_J^\ast=\frac{3}{7}(3+\sqrt{2})t_E^\ast~,\ee and \be \Omega(t_J)=\Big(\frac{-t_J}{-t_J^\ast}\Big)^{\frac{2+3\sqrt{2}}{7}}~,\ee where $t_J$ and $t_J^\ast$ still smaller than 0. The scale factor $a_J$, the Hubble parameter $H_J$ and the equation of state $w_J$ will be given by: \be \label{evolutionmb} a_J(t_J)\sim\Big(\frac{-t_J}{-t_J^\ast}\Big)^{\frac{4-\sqrt{2}}{7}}~,~H_J(t_J)=\frac{4-\sqrt{2}}{7t_J}~,~w_J=\frac{1+\sqrt{2}}{3}~. \ee

From this result we can see that, since the index of $a_J$ in terms of $t_J$ is larger than zero, so $a_J(t_J)$ will be decreasing as $t_J$ increases, indicating that there is also an contracting universe driven by $f(R)$ modified gravity theory when we require it be equivalent to matter-contraction scenario in GR when transformed to the Einstein frame. The Hubble parameter $H_J(t_J)$ is smaller than zero because of the negative $t_J$, and the equation of state $w_J$ of the universe is about the value of $0.8$, which is even larger.

\begin{figure}[htbp]
\centering
\includegraphics[scale=0.5]{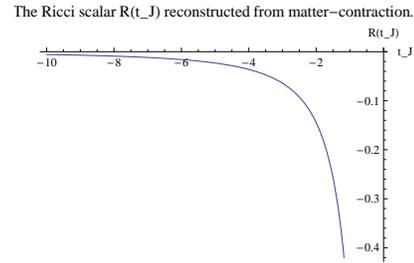}
\caption{The behavior of $R(t_J)$ w.r.t. $t_J$, where we choose $M=0.1$ and hence $t_E^\ast=10$. In this case, $R<0$, and is decreasing w.r.t. $t_J$.}\label{Ricci2plot}
\end{figure}

\begin{figure}[htbp]
\centering
\includegraphics[scale=0.5]{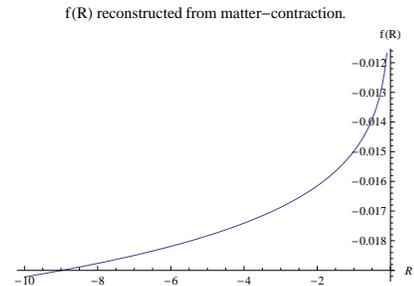}
\caption{The behavior of $f(R)$ w.r.t. $R$, where we choose $M=0.1$ and hence $t_E^\ast=10$. We can see that $f(R)$ is also less than 0, and increases with $R$ since both $R$ and $f(R)$ is decreasing w.r.t. $t_J$.}\label{f2}
\end{figure}

The Ricci scalar $R$ in this case is: \be\label{riccimb} R(t_J)=\frac{6(8-9\sqrt{2})}{49t_J^2}~,\ee which gives the form of $F(R)$ as: \be F(R)=\frac{1}{2}\Big(\frac{R}{R_0^{MC}}\Big)^{-\frac{2+3\sqrt{2}}{7}}~,~R_0^{MC}\equiv\frac{6(8-9\sqrt{2})}{49{t_J^\ast}^2}~,\ee and \bea\label{frmb} f(R)&=&\int F(R)dR~\nonumber\\ &=&\frac{5+3\sqrt{2}}{2}R_0^{MC}\Big(\frac{R}{R_0^{MC}}\Big)^{\frac{1}{(5+3\sqrt{2})}}~.\eea The plot of $R(t_J)$ and $f(R)$, which are reconstructed from matter-contraction in its Einstein frame, are presented in Figs. \ref{Ricci2plot} and \ref{f2}.


One can check our results with the conditions for generating scale-invariant power spectrum for consistency. For the case of inflation, from Eq. (\ref{evolutioninf}) we can write down the relation of conformal time $\eta$ and $t_J$ as: \bea \eta&=&\int a_J^{-1}(t_J)dt_J~\nonumber\\ &\sim&(-t_J)^{\frac{\sqrt{3}(1-\epsilon_E)}{\sqrt{\epsilon_E}(1-\sqrt{3\epsilon_E})}}~,\eea while \be a_J\sqrt{Q_{\cal R}}\sim a_J\sqrt{F}\sim (-t_J)^{\frac{\sqrt{3}}{\sqrt{\epsilon_E}(\sqrt{3\epsilon_E}-1)}}~,\ee where we note that $\delta_F$ is a constant. Thus we could easily find that \be a_J\sqrt{Q_{\cal R}}\sim\eta^{1/(\epsilon_E-1)}\sim\eta^{-1}~\ee when $|\epsilon_E|\ll1$. The case of matter-contraction is similar. From Eq. (\ref{evolutionmb}) one has: \be \eta\sim(-t_J)^{\frac{3+\sqrt{2}}{7}}~,\ee and \be a_J\sqrt{Q_{\cal R}}\sim a_J\sqrt{F}\sim (-t_J)^{\frac{2(3+\sqrt{2})}{7}}~,\ee which gives \be a_J\sqrt{Q_{\cal R}}\sim\eta^2~.\ee Moreover, one can also check the conditions for ghost-free and stable fluctuations for our constructed $f(R)$ models using the criterion for $f(R)$ models mentioned in e.g. Ref. \cite{Faraoni:2005ie}. From our expressions (\ref{frinf}) and (\ref{frmb}) one can easily check that the fluctuations in our models have neither ghost or instabilities.

Before ending this section, let's also remark the relation between the general evolutions of the universe driven by $f(R)$ modified gravity theory in the two frames with an arbitrary constant $\epsilon_E$, though without showing the detailed calculations. The relation between variables in the two frames is summarized in TABLE \ref{table}. Here we write $\Omega(t_E)$ in a general form of $\Omega=(\pm t_E/\pm t_E^\ast)^\omega$.
\begin{widetext}
\begin{table*}
\centering
\begin{tabular}{c|c|c|c|c|c|c|c}
\hline
\multirow{2}{*}{$t_E$}      & \multirow{2}{*}{$\epsilon_E$} & $a_E$ & $\omega$ & $t_J$ & $\epsilon_J$ & $a_J$ & $\rm{horizon}$ \\
& & $(\sim t_E^{1/\epsilon_E})$ & $(=1/\sqrt{3\epsilon_E})$ & $(\sim [t_E^\ast/(1-\omega)]t_E^{1-\omega})$ & $(=(\omega-1)/(\omega-1/\epsilon_E))$ & $(\sim t_J^{1/\epsilon_J})$ & $\rm{problem}$ \\ \hhline{========}
\multirow{4}{*}{$t_E>0$} & $\epsilon_E>3$     & \multirow{4}{*}{\rm{expanding}}   & $1/\epsilon_E<\omega<1$ & \multirow{3}{*}{$t_J>0$} & $\epsilon_J<0$   & \rm{contracting}                  & \multirow{2}{*}{y} \\
\cline{2-2} \cline{4-4} \cline{6-7}
                         & $1<\epsilon_E<3$   &                                   & $\omega<1/\epsilon_E$   &                          & $\epsilon_J>1$   & \multirow{3}{*}{\rm{expanding}}   &                    \\
\cline{2-2} \cline{4-4} \cline{6-6} \cline{8-8}
                         & $1/3<\epsilon_E<1$ &                                   & $\omega<1$              &                          & $0<\epsilon_J<1$ &                                   & \multirow{2}{*}{n} \\
\cline{2-2} \cline{4-6}
                         &$0<\epsilon_E<1/3$  &                                   &$1<\omega<1/\epsilon_E$  & $t_J<0$                  & $\epsilon_J<0$   &                                   &                    \\
\hline
\multirow{4}{*}{$t_E<0$} & $\epsilon_E>3$     & \multirow{4}{*}{\rm{contracting}} & $1/\epsilon_E<\omega<1$ & \multirow{3}{*}{$t_J<0$} & $\epsilon_J<0$   & \rm{expanding}                    & \multirow{2}{*}{n} \\
\cline{2-2} \cline{4-4} \cline{6-7}
                         & $1<\epsilon_E<3$   &                                   & $\omega<1/\epsilon_E$   &                          & $\epsilon_J>1$   & \multirow{3}{*}{\rm{contracting}} &                    \\
\cline{2-2} \cline{4-4} \cline{6-6} \cline{8-8}
                         & $1/3<\epsilon_E<1$ &                                   & $\omega<1$              &                          & $0<\epsilon_J<1$ &                                   & \multirow{2}{*}{y} \\
\cline{2-2} \cline{4-6}
                         & $0<\epsilon_E<1/3$ &                                   & $1<\omega<1/\epsilon_E$ & $t_J>0$                  & $\epsilon_J<0$   &                                   &                    \\
\hline
\end{tabular}
\caption{
The relations between variables in the Jordan and Einstein frames where we generalize $\epsilon_E$ to be an arbitrary positive constant value. $t_E$ can be chosen as either positive or negative, presenting parametrization of an expanding or a contracting universe. For $\epsilon_E>1$ in expanding phase or $\epsilon_E<1$ in contracting phase, we have horizon problem, while in the other two cases we don't. Due to the fact that $\omega=1/\sqrt{3\epsilon_E}$, the region of $t_J>0/<0$ can be divided by the line of $\omega=1(\epsilon_E=1/3)$, the region of $\epsilon_J>0/<0$ is divided by both $\omega=1$ and $\omega=1/\epsilon_E(\epsilon_E=1)$. In the $\epsilon_J>0$ region, the region of $\epsilon_J>1/<1$ is divided by the line of $\omega=1/\epsilon_E(\epsilon_E=3)$. Whether the universe contracts or expands in the Jordan frame is decided by whether $\epsilon_E>3$ or not. Finally, when there is no horizon problem in the Einstein frame, there will be no horizon problem in the Jordan frame, and vice versa. Similar summary but only for GR can be found in, e.g. \cite{Piao:2004jg}.}\label{table}
\end{table*}
\end{widetext}
Note that since $f(R)$ theory can only be equivalent to canonical field via conformal transformation, we don't have $\epsilon_E<0$ case.

\section{Discussions and Conclusion}
In this paper we studied the reconstruction and cosmic evolutions of $f(R)$ modified gravity models, which can be transformed as inflation or matter-contraction scenarios in their Einstein frame. The equivalence of the Jordan and Einstein frames guarantee that the perturbations generated by $f(R)$ models follows the same evolution, namely can give rise to scale-invariant power spectrum required by observations, however their background evolution might be different. In our previous work \cite{Qiu:2012ia} we have shown that there can be more than one kind of evolution in the case of general nonminimal coupling theories, but for the $f(R)$ case, there's no such degeneracy and the correspondence between the two frames must be one to one. We find that in $f(R)$ modified gravity theory, inflation in the Einstein frame can only refer to (phantom-like) inflation in the Jordan frame, while matter-contraction in the Einstein frame can only refer to contraction with a larger equation of state in the Jordan frame. We analysed the general conditions for $f(R)$ theory of getting scale-invariant power spectrum, and obtained the evolution of the universe in the Jordan frame as well as the functional form of $f(R)$. Numerical plot of $R$ w.r.t. $t_J$ and $f(R)$ w.r.t. $R$ are also presented.

In the current paper, we only focus on $f(R)$ models corresponds to models in Einstein frame with constant $\epsilon_E$. For case where $\epsilon_E$ is time-varying will also be interesting, and has been investigated in many places. Varying $\epsilon_E$ can also be one of the mechanisms of generating scale-invariant power spectrum, especially in scenarios alternative to inflation, see e.g. \cite{Khoury:2009my}. Moreover, for whole evolution process of the universe, including reheating after inflation or transfering to late-time acceleration. For these consideration, more complicated functional form of $f(R)$ models is needed. For example, for the reheating process, other field will be introduced to interact with gravity in order to produce particles effectively. This requires new conformal relations for multi-degrees of freedom other than Eq. (\ref{relationfr}). All these interesting topics are under investigation now.

Before ending, we would like to mention that due to the equivalence of the two frames, the Big-Bang cosmological problems (horizon, flatness, etc.) will also do no harm to the reconstructed $f(R)$ models. To see this, one can look into the efolding number $\cal N$ defined as \cite{Khoury:2003vb} \be {\cal N}\equiv\ln\Big(\frac{a_iH_i}{a_iH_i}\Big)~,\ee which can be directly related to these problems. Usually these problems can be avoided as long as we require that ${\cal N}\gtrsim 70$ during inflation. From the relation (\ref{relationfr}) we can see that the conformal Hubble parameter, ${\cal H}\equiv aH$, is not conformal invariant, but since in our case $\delta_F$ is a constant, ${\cal N}$ is a conformal invariant variable. Therefore, provided that inflation lasts for enough efolding number in one frame, one need not worry about whether it does in the other frame. We'd also like to refer the readers to \cite{Qiu:2012ia} for more detailed arguments.
\section*{Acknowledgments}
The author thanks Antonio de Felice, Je-An Gu and Yun-Song Piao for useful discussions. This work is funded in part by the National Science Council of R.O.C. under Grant No. NSC99-2112-M-033-005-MY3 and No. NSC99-2811-M-033-008 and by the National Center for Theoretical Sciences.


\begin{thebibliography}{99}

\bibitem{Qiu:2012ia}
  T.~Qiu,
  JCAP {\bf 1206}, 041 (2012)
  [arXiv:1204.0189 [hep-ph]].

\bibitem{Larson:2010gs}
  D.~Larson {\it et al.},
  Astrophys.\ J.\ Suppl.\  {\bf 192}, 16 (2011).

\bibitem{Bernardeau:2001qr}
  F.~Bernardeau, S.~Colombi, E.~Gaztanaga and R.~Scoccimarro,
  Phys.\ Rept.\  {\bf 367}, 1 (2002)
  [arXiv:astro-ph/0112551].

\bibitem{Starobinsky:1980te}
  A.~A.~Starobinsky,
  Phys.\ Lett.\ B {\bf 91}, 99 (1980).

\bibitem{Guth:1980zm}
  A.~H.~Guth,
  Phys.\ Rev.\  D {\bf 23}, 347 (1981).

\bibitem{Albrecht:1982wi}
  A.~Albrecht and P.~J.~Steinhardt,
  Phys.\ Rev.\ Lett.\  {\bf 48}, 1220 (1982); A.~D.~Linde,
  Phys.\ Lett.\  B {\bf 108}, 389 (1982).

\bibitem{Linde:1983gd}
  A.~D.~Linde,
  Phys.\ Lett.\  B {\bf 129} (1983) 177.

\bibitem{Starobinsky:1985ww}
  A.~A.~Starobinsky,
  Sov.\ Astron.\ Lett.\  {\bf 11} (1985) 133;
  A.~D.~Linde,
  Prog.\ Theor.\ Phys.\ Suppl.\  {\bf 163}, 295 (2006)
  [arXiv:hep-th/0503195];
  K.~A.~Olive,
  Phys.\ Rept.\  {\bf 190}, 307 (1990);
  D.~Boyanovsky, C.~Destri, H.~J.~De Vega and N.~G.~Sanchez,
  Int.\ J.\ Mod.\ Phys.\  A {\bf 24}, 3669 (2009)
  [arXiv:0901.0549 [astro-ph.CO]];
  A.~Mazumdar and J.~Rocher,
  Phys.\ Rept.\  {\bf 497}, 85 (2011)
  [arXiv:1001.0993 [hep-ph]].

\bibitem{Finelli:2001sr}
  F.~Finelli and R.~Brandenberger,
  Phys.\ Rev.\  D {\bf 65}, 103522 (2002)
  [arXiv:hep-th/0112249].

\bibitem{Cai:2007qw}
  Y.~F.~Cai, T.~Qiu, Y.~S.~Piao, M.~Li and X.~Zhang,
  JHEP {\bf 0710}, 071 (2007)
  [arXiv:0704.1090 [gr-qc]];
  Y.~F.~Cai, T.~t.~Qiu, R.~Brandenberger and X.~m.~Zhang,
  Phys.\ Rev.\  D {\bf 80}, 023511 (2009)
  [arXiv:0810.4677 [hep-th]];
  T.~Qiu and K.~C.~Yang,
  JCAP {\bf 1011}, 012 (2010)
  [arXiv:1007.2571 [astro-ph.CO]];
  J.~Karouby, T.~Qiu and R.~Brandenberger,
  Phys.\ Rev.\ D {\bf 84}, 043505 (2011)  [arXiv:1104.3193 [hep-th]].

\bibitem{Faraoni:1998qx}
  V.~Faraoni, E.~Gunzig and P.~Nardone,
  Fund.\ Cosmic Phys.\  {\bf 20}, 121 (1999)
  [arXiv:gr-qc/9811047].

\bibitem{Faraoni:2000gx}
  V.~Faraoni,
  Int.\ J.\ Theor.\ Phys.\  {\bf 40}, 2259 (2001)
  [arXiv:hep-th/0009053];
  A.~De Felice and S.~Tsujikawa,
  Living Rev.\ Rel.\  {\bf 13}, 3 (2010)
  [arXiv:1002.4928 [gr-qc]];
  S.~Nojiri and S.~D.~Odintsov,
  Phys.\ Rept.\  {\bf 505}, 59 (2011)
  [arXiv:1011.0544 [gr-qc]].

\bibitem{Nojiri:2006be}
  S.~Nojiri and S.~D.~Odintsov,
  J.\ Phys.\ Conf.\ Ser.\  {\bf 66}, 012005 (2007)
  [arXiv:hep-th/0611071];
  X.~Wu and Z.~H.~Zhu,
  Phys.\ Lett.\  B {\bf 660}, 293 (2008)
  [arXiv:0712.3603 [astro-ph]];
  E.~Elizalde, S.~Nojiri, S.~D.~Odintsov, D.~Saez-Gomez and V.~Faraoni,
  Phys.\ Rev.\  D {\bf 77}, 106005 (2008)
  [arXiv:0803.1311 [hep-th]];
  L.~N.~Granda,
  arXiv:0812.1596 [hep-th];
  K.~Bamba, C.~Q.~Geng, S.~Nojiri and S.~D.~Odintsov,
  Phys.\ Rev.\  D {\bf 79}, 083014 (2009)
  [arXiv:0810.4296 [hep-th]];
  S.~Nojiri, S.~D.~Odintsov and D.~Saez-Gomez,
  Phys.\ Lett.\  B {\bf 681}, 74 (2009)
  [arXiv:0908.1269 [hep-th]];
  S.~Nojiri, S.~D.~Odintsov, A.~Toporensky and P.~Tretyakov,
  Gen.\ Rel.\ Grav.\  {\bf 42}, 1997 (2010)
  [arXiv:0912.2488 [hep-th]];
  J.~H.~He, B.~Wang and E.~Abdalla,
  Phys.\ Rev.\  D {\bf 83}, 063515 (2011)
  [arXiv:1012.3904 [astro-ph.CO]];
  J.~h.~He and B.~Wang,
  arXiv:1203.2766 [astro-ph.CO];
  arXiv:1208.1388 [astro-ph.CO].

\bibitem{Arnowitt:1962hi}
  R.~L.~Arnowitt, S.~Deser and C.~W.~Misner,
  arXiv:gr-qc/0405109.

\bibitem{Gunzig:2000kk}
  E.~Gunzig, A.~Saa, L.~Brenig, V.~Faraoni, T.~M.~Rocha Filho and A.~Figueiredo,
  Phys.\ Rev.\  D {\bf 63}, 067301 (2001)
  [arXiv:gr-qc/0012085];
  A.~Saa, E.~Gunzig, L.~Brenig, V.~Faraoni, T.~M.~Rocha Filho and A.~Figueiredo,
  Int.\ J.\ Theor.\ Phys.\  {\bf 40}, 2295 (2001)
  [arXiv:gr-qc/0012105];
  M.~Baldi, F.~Finelli and S.~Matarrese,
  Phys.\ Rev.\  D {\bf 72}, 083504 (2005)
  [arXiv:astro-ph/0505552].

\bibitem{Piao:2003ty}
  Y.~S.~Piao and E.~Zhou,
  Phys.\ Rev.\  D {\bf 68}, 083515 (2003)
  [arXiv:hep-th/0308080];
  Y.~S.~Piao and Y.~Z.~Zhang,
  Phys.\ Rev.\  D {\bf 70}, 063513 (2004)
  [arXiv:astro-ph/0401231];
  J.~E.~Lidsey,
  Phys.\ Rev.\  D {\bf 70}, 041302 (2004)
  [arXiv:gr-qc/0405055];
  P.~F.~Gonzalez-Diaz and J.~A.~Jimenez-Madrid,
  Phys.\ Lett.\  B {\bf 596}, 16 (2004)
  [arXiv:hep-th/0406261];
  S.~Capozziello, S.~Nojiri and S.~D.~Odintsov,
  Phys.\ Lett.\  B {\bf 632}, 597 (2006)
  [arXiv:hep-th/0507182];
  S.~Nojiri and S.~D.~Odintsov,
  Gen.\ Rel.\ Grav.\  {\bf 38}, 1285 (2006)
  [arXiv:hep-th/0506212];
  P.~Wu and H.~W.~Yu,
  JCAP {\bf 0605}, 008 (2006)
  [arXiv:gr-qc/0604117];
  C.~J.~Feng, X.~Z.~Li and E.~N.~Saridakis,
  Phys.\ Rev.\  D {\bf 82}, 023526 (2010)
  [arXiv:1004.1874 [astro-ph.CO]];
  Y.~S.~Piao,
  Phys.\ Rev.\  D {\bf 78}, 023518 (2008)
  [arXiv:0712.3328 [gr-qc]];
  Z.~G.~Liu, J.~Zhang and Y.~S.~Piao,
  Phys.\ Lett.\  B {\bf 697}, 407 (2011)
  [arXiv:1012.0673 [gr-qc]];
  Z.~G.~Liu and Y.~S.~Piao,
  arXiv:1203.4901 [gr-qc].
  
\bibitem{Faraoni:2005ie}
  V.~Faraoni,
  Phys.\ Rev.\ D {\bf 72}, 061501 (2005)  
  [gr-qc/0509008];
  V.~Faraoni,
  Phys.\ Rev.\ D {\bf 72}, 061501 (2005)  
  [gr-qc/0509008];
  V.~Faraoni and S.~Nadeau,
  Phys.\ Rev.\ D {\bf 75}, 023501 (2007)  
  [gr-qc/0612075];
  V.~Faraoni,
  Phys.\ Rev.\ D {\bf 75}, 067302 (2007)  
  [gr-qc/0703044 [GR-QC]].

\bibitem{Piao:2004jg}
  Y.~-S.~Piao and Y.~-Z.~Zhang,
  Phys.\ Rev.\ D {\bf 70}, 043516 (2004)
  [astro-ph/0403671];
  Y.~-S.~Piao,
  Phys.\ Lett.\ B {\bf 606}, 245 (2005)
  [hep-th/0404002].

\bibitem{Khoury:2009my}
  J.~Khoury and P.~J.~Steinhardt,
  Phys.\ Rev.\ Lett.\  {\bf 104}, 091301 (2010)  
  [arXiv:0910.2230 [hep-th]];
  A.~Linde, V.~Mukhanov and A.~Vikman,
  JCAP {\bf 1002}, 006 (2010)  
  [arXiv:0912.0944 [hep-th]];
  Y.~-S.~Piao,
  Phys.\ Lett.\ B {\bf 701}, 526 (2011)  
  [arXiv:1012.2734 [hep-th]];
  J.~Khoury and P.~J.~Steinhardt,
  Phys.\ Rev.\ D {\bf 83}, 123502 (2011)  
  [arXiv:1101.3548 [hep-th]].
  
\bibitem{Khoury:2003vb}
  J.~Khoury, P.~J.~Steinhardt and N.~Turok,
  Phys.\ Rev.\ Lett.\  {\bf 91}, 161301 (2003)
  [astro-ph/0302012].

\end{thebibliography}
\end{document}